\documentclass[preprint,aps,prd,showpacs,nofootinbib,tightenlines]{revtex4-1}
\usepackage{amsmath}
\usepackage{bm}
\usepackage{times}
\usepackage{braket}
\usepackage{epsfig}
\usepackage{color,graphicx}
\usepackage{slashed}
\usepackage{hyperref}
\definecolor{Red}{rgb}{1.,0.,0.}

\definecolor{Blue}{rgb}{0.,0.,1.}

\definecolor{nicered}{rgb}{0.7,0.1,0.1}
\definecolor{nicegreen}{rgb}{0.1,0.5,0.1}

\hypersetup{colorlinks,citecolor=nicegreen,linkcolor=nicered}

\begin{document}

\newcommand{\beq}{\begin{eqnarray}}
\newcommand{\eeq}{\end{eqnarray}}

\newcommand{\non}{\nonumber\\ }
\newcommand{\ov}{\overline}
\newcommand{\rmt}{ {\rm T}}
\newcommand{\psl}{ p \hspace{-2.0truemm}/ }
\newcommand{\qsl}{ q \hspace{-2.0truemm}/ }
\newcommand{\epsl}{ \epsilon \hspace{-2.0truemm}/ }
\newcommand{\nsl}{ n \hspace{-2.2truemm}/ }
\newcommand{\vsl}{ v \hspace{-2.2truemm}/ }

\newcommand{\bsb}{ \bar{B}_s^0}
\newcommand{\etar}{ \eta^\prime }
\newcommand{\etap}{ \eta^{(\prime)} }
\newcommand{\jpsi}{ J/\Psi }

\def \cpc{ {\bf Chin. Phys. C } }
\def \ctp{ {\bf Commun. Theor. Phys. } }
\def \epjc{{\bf Eur. Phys. J. C} }
\def \jhep{{\bf JHEP } }
\def \jpg{ {\bf J. Phys. G} }
\def \mpla{ {\bf Mod. Phys. Lett. A } }
\def \npb{ {\bf Nucl. Phys. B} }
\def \plb{ {\bf Phys. Lett. B} }
\def \pr{ {\bf Phys. Rep.} }
\def \prd{ {\bf Phys. Rev. D} }
\def \prl{ {\bf Phys. Rev. Lett.}  }
\def \ptp{ {\bf Prog. Theor. Phys. }  }
\def \zpc{ {\bf Z. Phys. C}  }


\title{Study of $\bar{B}_s^0 \to (D_s^+,D_s^{*+}) l^-\bar{\nu}_l$ decays in the pQCD factorization approach}
\author{Ying-Ying Fan$^{1}$, Wen-Fei Wang$^{2}$, and Zhen-Jun Xiao$^{1}$\footnote{xiaozhenjun@njnu.edu.cn} }
\affiliation{1. Department of Physics and Institute of Theoretical Physics,\\
Nanjing Normal University, Nanjing, Jiangsu 210023, P.R. China,\\
2. Institute of High Energy Physics and Theoretical Physics Center
for Science Facilities,\\
 CAS, P.O.Box 918(4), Beijing 100049, P.R. China }

\date{\today}
\begin{abstract}
The $\bar{B}_s^0 \to D_s^{(*)+} l^- \bar{\nu}_l$ semileptonic decays
were calculated in the framework of the standard model (SM) by employing
the perturbative QCD (pQCD) factorization approach. We defined
four ratios of the branching ratios of the considered decays $R(D_s^{(*)})$ and $R_{D_s}^{l,\tau}$.
From the numerical results and phenomenological analysis we found that:
(a) The pQCD predictions for the branching ratios $Br(\bar{B}_s^0 \to D_s^{(*)+}
l^-\bar{\nu}_l)$ generally agree well with the previous theoretical predictions;
(b) For the four ratios, the pQCD predictions are $R(D_s)= 0.392 \pm 0.022$,
$R(D_s^{*})= 0.302\pm 0.011$, $R_{D_{s}}^l=0.448^{+0.058}_{-0.041}$ and $R_{D_{s}}^\tau=0.582^{+0.071}_{-0.045}$,
which show a very good $SU(3)_F$ flavor symmetry with the corresponding ratios for
$B\to D^{(*)} l^- \bar{\nu}_l$ decays;
and (c) we strongly suggest the measurements of the new ratios $R(D_s^{(*)})$  and $R_{D_{s}}^{l,\tau}$
in the forthcoming Super-B experiments.
\end{abstract}


\maketitle

\section{Introduction}\label{sec:1} 

In Ref.~\cite{prl109-101802}, the BaBar collaboration reported a
combined $3.4\sigma$ deviation of their measured ratios $R(D)$ and
$R(D^*)$ from the standard model (SM) predictions. The measured
values  are \cite{prl109-101802}
\beq
{\cal R}(D)= 0.440 \pm 0.072, \quad
{\cal R}(D^*)= 0.332 \pm 0.030,\label{eq:exp01}
\eeq
while the SM predictions obtained by using the traditional
heavy quark effective theory (HQET) \cite{hqet1,hqet2} to evaluate
the form factors of $B \to D, D^*$ transitions are  the following  \cite{prd85-094025}:
\beq
{\cal R}(D)^{SM} = 0.296(16), \quad {\cal R}(D^*)^{SM} = 0.252(3).
\eeq
which are indeed much smaller than those measured values as given in Eq.~(1).
This $R(D^{(*)})$ anomaly has invoked intensive studies about the semileptonic $B \to D^{(*)} l^- \bar{\nu}_l$
decays\cite{prl109-071802,prl109-161801,prd86-054014, jhep2013-01054,prd86-034027,prd86-114037,
mpla27-1250183,Kosnik-1301} in the framework of the SM by employing
the different mechanisms or methods, but they all failed to interpret the data.

Motivated by the great difference between the theoretical predictions and the BaBar's
measurements about the ratios $R(D^*)$, we calculated the ratios $R(D^*)$ by
employing the perturbative QCD (pQCD) factorization approach \cite{li2003}
to evaluate the $B \to (D, D^*)$ transition form factors, and then found
numerically that \cite{fan2013a}
\beq
R(D)=0.430^{+0.021}_{-0.026}, \quad R(D^*)=0.301\pm 0.013.\label{eq:pqcd1}
\eeq
These pQCD predictions agree very well with the BaBar's measurements.

Among the various $B/B_s$ semileptonic decays, the $\bar{B}_s^0 \to
D_s^{(*)+} l^- \bar{\nu}_l$ decays are closely related with those
$B \to D^{(*)} l^- \bar{\nu}_l$ decays through the
$SU(3)_F$ flavor symmetry: they are all controlled by the same $b \to c
l^-\bar{\nu}_l$ transitions at the quark level, but with a different
spectator quark, from the $\bar{s}$ quark to the $\bar{u}$ or $\bar{d}$ quark: i.e.
 \beq
\underbrace{\bar{B}^0_s \to D_s^{(*)+} l^- \bar{\nu}_l}_{spectator \
\ is \ \ \bar{s}} \Longleftrightarrow \underbrace{B^-/\bar{B}^0 \to
D^{(*)} l^- \bar{\nu}_l}_{spectator \ \ is \ \ (\bar{u},\bar{d})}
\label{eq:001}
 \eeq
In the limit of $SU(3)$ flavor symmetry, these two kinds of decays should have
very similar properties. It is therefore very interesting to make a systematic
study for the $\bar{B}_s^0 \to D_s^{(*)+} l^- \bar{\nu}_l$ decays, and most
importantly  measure them in the forthcoming Super-B experiments, even if
the LHCb can not do the job due to the escape of the neutrinos.

In this paper, we will study $\bar{B}^0_s \to  D_s^{(*)+} l^- \bar{\nu}_l $ decays by employing
the pQCD factorization approach. Analogous to $R(D^{(*)})$ for $B \to D^{(*)}$ transitions, we here  define the similar
ratios of the branching ratios in the form of:
\beq
{\cal R}(D_s) \equiv  \frac{{\cal B}(\bar{B}^0_s \to D_s^+ \tau^- \bar{\nu}_\tau)}{{\cal B}
(\bar{B}_s^0 \to D_s^+ l^- \bar{\nu}_l)}, \qquad
{\cal R}(D_s^*)  \equiv  \frac{{\cal B}(\bsb \to D_s^{*+} \tau^- \bar{\nu}_\tau)}{{\cal B}(
\bsb \to D_s^{*+} l^- \bar{\nu}_l)}, \label{eq:rnew1}
\eeq
where $l^-=(e^-,\mu^-)$, which measures the mass effects of the heavy $\tau$ and light $e^-$ or $\mu^-$ leptons.
Following Ref.~\cite{fan2013a}, furthermore, we here also define other two ratios in the form of
\beq
{\cal R}^l_{D_s} \equiv  \frac{ \sum_{l=e,\mu}{\cal B}(\bar{B}^0_s \to D_s^+ l^- \bar{\nu}_l)}
{\sum_{l=e,\mu}{\cal B} (\bar{B}_s^0 \to D_s^{*+} l^- \bar{\nu}_l)}, \qquad
{\cal R}^\tau_{D_s}  \equiv  \frac{{\cal B}(\bsb \to D_s^{+} \tau^- \bar{\nu}_\tau)}{{\cal B}(
\bsb \to D_s^{*+} \tau^- \bar{\nu}_\tau)}.\label{eq:rnew2}
\eeq
It is easy to see that these two ratios reveals the effects induced by the different
form factors  of $B_s \to D_s$ and $B_s\to D_s^*$ transitions, and
can also be measured in the future Super-B experiments.

Theoretically, the  semileptonic $B_s \to (D_s,D_s^*) l\bar{\nu}$ decays have been studied
frequently in the frame work of the SM.
The branching ratios of these decay modes have been studied, for example, in terms of the constituent
quark meson (CQM) model \cite{epjc51-601},
in the framework of the QCD sum rules (QCDSRs) \cite{prd78-054011} , in the light cone sum rules (LCSRs)
or the covariant light-front quark model (CLFQM) \cite{prd80-014005,prd82-094031}.
In Ref.~\cite{jpg39-045002,prd87-034033}, $B_s^0 \to (D_s^-, D_s^{*-}) $ transition form factors are
estimated by using the method based on  an instantaneous approximated Mandelstam formulation (IAMF) and the
instantaneous Bethe-Salpeter equation, or the relativistic quark model. The numerical predictions
as presented in all these mentioned works \cite{epjc51-601,prd78-054011,prd80-014005,prd82-094031,jpg39-045002,prd87-034033}
will be listed in Table III of this paper for the purpose of comparisons.

On the experiments side, the  semileptonic $B_s \to (D_s,D_s^*) l\bar{\nu}$
decays have not been  measured up to now.  In LHCb experiments, it can not be measured too
since the neutrino is inaccessible there. But in the forthcoming Super-B experiments,
these semileptonic $B_s$ decays with a neutrino as one of the final state lepton can be measured
precisely.

\begin{figure}[tbp]
\vspace{-4cm} \centerline{\epsfxsize=16cm \epsffile{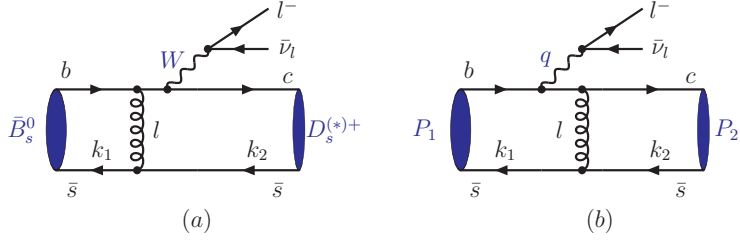}}
\vspace{-15.5cm} \caption{ The typical Feynman diagrams for the
semileptonic decays $\bar{B}^0_s \to (D_s^+,D_s^{*+}) l^-\bar{\nu}_l$ in the pQCD approach.}
\label{fig:fig1}
\end{figure}

In the pQCD factorization approach, the lowest order Feynman diagrams for $\bar{B}_s^0 \to D_s^{(*)+}
l^- \bar{\nu}_l$ decays are displayed in Fig.\ref{fig:fig1}, where the leptonic pairs come from the
$b$-quark's weak decay through charged current. The study for the
semileptonic decays $\bar{B}_s^0 \to (D^+_s,D_s^{*+}) l^-
\bar{\nu}_l$ can certainly be a great help for us to understand the BaBar's measurements for $R(D^{(*)})$.

The paper is organized as follows: In Sec.~\ref{sec:2}, we firstly
give a short review for the kinematics of the $\bar{B}_s^0 \to
D_s^{(*)+} l^- \bar{\nu}_l$ decays, and then we make a pQCD
calculation for the form factors $F_{0,+}(q^2)$, $V(q^2)$ and
$A_{0,1,2}(q^2)$ for $\bar{B}_s^0 \to D_s^{(*)+}$ transitions, and
present the formulaes for the differential decay rates of the
considered decay modes. In Sec.~III, we will present the pQCD
predictions for the branching ratios of all considered decays, as
well as the ratios $R(D_s^{(*)})$ and $R_{D_{s}}^{l,\tau}$ and make a comparative study with those currently
known theoretical predictions. The final section contains the conclusions and a short summary.

\section{Theoretical framework}\label{sec:2}

\subsection{Kinematics and the wave functions}\label{sub:21}

In the $B_s$ meson rest frame, we define the $B_s$ meson momentum
$P_1$, the $D_s^{(*)}$ momentum $P_2$ and the polarization vectors
$\epsilon$ of the $D^{*}_s$ in the light-cone coordinates as\cite{prd67-054028}
 \beq\label{eq-mom-p1p2}
P_1&=&\frac{m_{B_s}}{\sqrt{2}}(1,1,0_\bot),\quad P_2=\frac{r
m_{B_s}}{\sqrt{2}} (\eta^+,\eta^-,0_\bot),\non
\epsilon_L&=&\frac{1}{\sqrt{2}}(\eta^+,-\eta^-,0_\bot), \quad
\epsilon_T=(0,0,1),
 \eeq
  with the ratio $r=m_{D_s}/m_{B_s}$ or
$r=m_{D^{*}_s}/m_{B_s}$. The factors $\eta^\pm = \eta \pm
\sqrt{\eta^2-1}$ is defined in terms of the parameter $\eta =
\frac{1}{2r}\left [ 1+r^2-\frac{q^2}{m_{B_s}^2}\right]$  as in
Ref.~\cite{prd67-054028}, while $\epsilon_L$ and $\epsilon_T$
denotes the longitudinal and transverse polarization of the $D_s^*$
meson, respectively. The momenta of the spectator quarks in
$\bar{B}^0_s$ and $D_s^{(*)}$ mesons are parameterized as
 \beq
\label{eq:k1k2} k_1 =(0,x_1\frac{m_{B_s}}{\sqrt{2}},k_{1\bot}),
\quad k_2=\frac{m_{B_s}}{\sqrt{2}}(x_2r\eta^+,x_2r\eta^-,k_{2\bot}).
\eeq

For the $B_s$ meson wave function, we make use of the same
parameterizations as in Refs.~\cite{prd85-094003,prd86-114025} and
we adopt the B-meson distribution amplitude widely used in the pQCD
approach
 \beq \phi_{B_s}(x,b)&=& N_{B_s} x^2(1-x)^2\mathrm{\exp}
\left
 [ -\frac{M_{B_s}^2\; x^2}{2 \omega_{B_s}^2} -\frac{1}{2} (\omega_{B_s} b)^2\right],
\label{eq:phib} \eeq
 where the normalization factor $N_{B_s}$
depends on the values of the parameter $\omega_{B_s}$ and decay
constant $f_{B_s}$ and defined through the normalization relation
\cite{prd86-114025}. For $\omega_{B_s}$, one usually take
$\omega_{B_s}=0.50\pm 0.05$ GeV for $B_s^0$ meson.

For the pseudoscalar meson $D_s$, the wave function is in the form of
\beq
\label{eq:phids}
\Phi_{D_s}(p_2,x)=\frac{i}{\sqrt{6}}\gamma_5 (\psl_2 + m_{D_s})\phi_{D_s}(x).
\eeq
For the vector $D^{*}_s$ meson, we take the wave function as
follows,
\beq
\Phi_{D_s^*}(p_2,x) = \frac{-i}{\sqrt{6}} \left
 [  \epsl_L(\psl_2  + m_{ D^*_s})\phi^L_{D^*_s}(x)
+ \epsl_T(\psl_2  + m_{D^*_s})\phi^T_{D^*_s}(x)\right ] .
\eeq
For the distribution amplitudes of $D_s$ and $D_s^*$ meson, we adopt the same one
as defined in Ref.~\cite{prd78-014018}
\beq
\phi_{ D_s }(x)=\frac{ 3 f_{D_s} }{\sqrt{6}} \;x(1-x)
\left[ 1+ C_{D_s}  (1-2x)\right]\exp \left[-\frac{\omega_{D_s}^2 b^2 }{2}\right],
\label{eq:phid}\\
\phi_{D_s^*}(x)=\frac{3 f_{D_s^{*} }}{\sqrt{6}}\; x(1-x)
\left[ 1+ C_{D_s^{*} } (1-2x)\right]\exp \left[-\frac{\omega_{D_s^*}^2 b^2 }{2}\right].
\label{eq:phids2}
\eeq
From the heavy quark limit, we here assume that
\beq
f^L_{D_s^*}=f^T_{D_s^*}=f_{D_s^*}, \quad
\phi^L_{D_s^*}=\phi^T_{D_s^*}=\phi_{D_s}.\label{eq:app01}
\eeq
For the $D_s^{(*)}$ mesons we also set $C_{D_s}=C_{D_s^*}=0.5,
\omega_{D_s}=\omega_{D_s^*} =0.1$ GeV \cite{prd78-014018} in the calculations.
Of course, the three distribution amplitudes $\phi^L_{D^*}$, $\phi^T_{D^*}$
and $\phi_{D^*}$ should be different according to the general expectations,
but we currently do not have other possible choices. The approximations as given
in Eq.~(\ref{eq:app01}) may be too simple, even poor.


\subsection{Form factors and differential decay rates}\label{sec:3b}

The form factors of the $\bar{B}_s^0\to D_s$ transition induced
by vector currents are defined as\cite{prd86-114025}:
 \beq \langle
D_s(p_2)|\bar{c}(0)\gamma_{\mu}b(0)|\bar{B}^0_s(p_1)\rangle&=&
F_+(q^2)
\left[(p_1+p_2)_{\mu}-\frac{m_{B_s}^2-m_{D_s}^2}{q^2}q_{\mu}\right]
\non && + F_0(q^2) \frac{m_{B_s}^2-m_{D_s}^2}{q^2}q_{\mu},\ \
 \eeq
where $q=p_1-p_2$ is the lepton-pair momentum. In order to cancel
the poles at $q^2=0$, $F_+(0)$ should be equal to $F_0(0)$. For the
sake of the calculation, it is convenient to define the auxiliary
form factors $f_1(q^2)$ and $f_2(q^2)$,
 \beq \langle
D_s(p_2)|\bar{c}(0)\gamma_{\mu}b(0)|\bar{B}^0_s(p_1)\rangle=f_1(q^2)p_{1\mu}+f_2(q^2)p_{2\mu},
\eeq
 where the form factors $f_1(q^2)$ and $f_2(q^2)$ are related to
$F_+(q^2)$ and $F_0(q^2)$ through the relation,
 \beq \label{eq:fpf0}
F_+(q^2)&=&\frac12\left[f_1(q^2)+f_2(q^2)\right], \notag\\
F_0(q^2)&=&\frac12 f_1(q^2)\left[1+\frac{q^2}{m_{B_s}^2-m_{D_s}^2}\right]
+\frac12 f_2(q^2)\left[1-\frac{q^2}{m_{B_s}^2-m_{D_s}^2}\right].
\eeq

For the $\bar{B}_s^0 \to D_s^+ l^- \bar{\nu}_l$ decays, by
analytical calculations in the pQCD approach we find the
$\bar{B}_s^0\to D_s$ form factors $f_1(q^2)$ and $f_2(q^2)$ as
following:
\beq  
f_1(q^2)&=&8\pi m^2_{B_s} C_F\int dx_1 dx_2\int b_1 db_1 b_2
db_2 \phi_{B_s}(x_1,b_1) \phi_{D_s}(x_2,b_2)\non &\times
&\Bigl\{\left[ 2 r\left(1-rx_2\right)
 \right]\cdot h_1(x_1,x_2,b_1,b_2)\cdot \alpha_s(t_1)
\cdot  \exp\left [-S_{ab}(t_1) \right ] \non
&+& \left[2 r(2 r_c-r)+x_1 r (-2+2 \eta+\sqrt{\eta^2-1}-\frac{2
\eta}{\sqrt{\eta^2-1}}+\frac{\eta^2}{\sqrt{\eta^2-1}}) \right]\non
 &\cdot& h_2(x_1,x_2,b_1,b_2)
\cdot \alpha_s (t_2)\cdot \exp\left [-S_{ab}(t_2) \right] \Bigr \},
\label{eq:f1q2}
\eeq
\beq
f_2(q^2)&=&8\pi m^2_{B_s} C_F\int dx_1 dx_2\int b_1 db_1 b_2 db_2
\phi_{B_s}(x_1,b_1)\phi_{D_s}(x_2,b_2)\non
&& \hspace{-1cm}\times  \Bigl\{ \left[ 2-4 x_2 r(1-\eta) \right] \cdot h_1(x_1,x_2,b_1,b_2)\cdot
\alpha_s(t_1)\cdot  \exp\left [-S_{ab}(t_1) \right ] \non
&&\hspace{-1cm}+ \left[ 4r-2r_c-x_1+\frac{x_1}{\sqrt{\eta^2-1}}(2-\eta) \right]\cdot
h_2(x_1,x_2,b_1,b_2) \cdot \alpha_s (t_2)\cdot \exp\left [-S_{ab}(t_2)
\right] \Bigr \},
\label{eq:f2q2}
\eeq 
where $C_F=4/3$ is a color factor, $r=m_{D_s}/m_{B_s}$, $r_c=m_c/m_{D_s}$ with the $m_c$ is the mass of $c$-quark.
The hard functions $h(x_i,b_i)$, the hard scales $t_{1,2}$ and the Sudakov factors $S_{ab}$ will be
given in the Appendix~\ref{sec:app1}.

For the charged current $\bar{B}_s^0\to D_s^+ l^-\bar{\nu}_l$
decays, the quark level transitions are the $b\to c l^-\bar{\nu}_l$
with the effective Hamiltonian:
 \beq \label{eq-hamiltonian} {\cal
H}_{eff}(b\to c l\bar \nu_l)=\frac{G_F}{\sqrt{2}}V_{cb}\; \bar{c}
\gamma_{\mu}(1-\gamma_5)b \cdot \bar l\gamma^{\mu}(1-\gamma_5)\nu_l,
\eeq
 where $G_F=1.166 37\times10^{-5} GeV^{-2}$ is the
Fermi-coupling constant. With the two form factors $F_+(q^2)$ and
$F_0(q^2)$, we can write down the differential decay rate of the
decay mode $\bar{B}_s^0 \to D_s^+ l^-\bar{\nu}_l$
as\cite{prd79-014013}:
 \beq
\frac{d\Gamma(b \to c
l\bar{\nu}_l)}{dq^2}&=&\frac{G_F^2|V_{cb}|^2}{192 \pi^3  m_{B_s}^3}
\left ( 1-\frac{m_l^2}{q^2} \right)^2\frac{
\lambda^{1/2}(q^2)}{2q^2}\cdot \Bigl \{  3 m_l^2\left
(m_{B_s}^2-m_{D_s}^2 \right )^2 |F_0(q^2)|^2 \non && + \left
(m_l^2+2q^2 \right )\lambda(q^2)|F_+(q^2)|^2 \Bigr \},
\label{eq:dg1}
 \eeq
  where $m_l$ is the mass of the leptons $e^-$,
$\mu^-$ or $\tau^-$. For the cases of $l^-=(e^-,\mu^-)$, the
corresponding mass terms $m_l^2$ could be neglected, the
Eq.~(\ref{eq:dg1}) then becomes very simple,
 \beq \frac{d\Gamma(b
\to c l\bar{\nu}_l)}{dq^2}&=&\frac{G_F^2}{192 \pi^3  m_{B_s}^3}
\lambda^{3/2}(q^2)|V_{cb}|^2 |F_+(q^2)|^2, \label{eq:dg1b}
 \eeq
where $\lambda(q^2)=(m_{B_s}^2+m_{D_s}^2-q^2)^2-4m_{B_s}^2m_{D_s}^2$
is the phase space factor.

For the $\bar{B}_s^0 \to D_s^*$ transitions, the hadronic matrix
elements of the vector and axial-vector currents are described by
the four QCD form factors $V(q^2)$ and $A_{0,1,2}(q^2)$  via~\cite{prd65-014007}:
\beq
\langle
D_s(p_2,\epsilon^*)|\bar{c}(0)\gamma_{\mu}b(0)|\bar{B}^0_s(p_1)\rangle
&=& \frac{2 i V(q^2)}{m_{B_s}+m_{D_s^*}}\epsilon_{\mu \nu \alpha
\beta} \epsilon^{* \nu}p_1^\alpha p_2^\beta,\non \langle
D_s(p_2,\epsilon^*)|\bar{c}(0)\gamma_{\mu}\gamma_5
b(0)|\bar{B}^0_s(p_1)\rangle &=& 2 m_{D_s^*}
A_0(q^2)\frac{\epsilon^*\cdot q}{q^2}q_\mu \non
&& + (m_{B_s} +
m_{D_s^*})A_1(q^2) \left (\epsilon^*_\mu - \frac{\epsilon^*\cdot q}{q^2}q_\mu \right )\non
& & \hspace{-1cm} - A_2(q^2)\frac{\epsilon^*\cdot
q}{m_{B_s} + m_{D_s^*}} \left [(p_1+p_2)_\mu -
\frac{m_{B_s}^2-m_{D_s^*}^2}{q^2}q_\mu \right ].
 \eeq

In the pQCD approach, we find the form factors $V(q^2)$ and
$A_{0,1,2}(q^2)$ for $\bar{B}_s^0\to D^*_s$ transition are of the
form:
\beq  
V(q^2)&=&8\pi m^2_{B_s} C_F\int dx_1 dx_2\int b_1 db_1 b_2 db_2
\phi_{B_s}(x_1,b_1)\phi^T_{D^*_s}(x_2,b_2) \cdot (1+r)\non
&\times & \Bigl \{\left[1-rx_2\right] \cdot h_1(x_1,x_2,b_1,b_2)\cdot
\alpha_s(t_1) \cdot \exp\left [-S_{ab}(t_1) \right ] \non &+&
\left[r+\frac{x_1}{2\sqrt{\eta^2-1}}\right] \cdot
h_2(x_1,x_2,b_1,b_2) \cdot \alpha_s (t_2)\cdot \exp\left
[-S_{ab}(t_2) \right] \Bigr \}, \label{eq:Vqq}
\eeq
\beq
A_0(q^2)&=&8\pi m^2_{B_s} C_F\int dx_1 dx_2\int b_1 db_1 b_2 db_2
\phi_{B_s}(x_1,b_1)\phi^L_{D^*_s}(x_2,b_2)\non &\times & \Bigl \{
\left[ 1+r -rx_2(2+r-2\eta)\right]\cdot h_1(x_1,x_2,b_1,b_2)\cdot
\alpha_s(t_1) \cdot \exp\left [-S_{ab}(t_1) \right ] \non &+& \left
[r^2+r_c+\frac{x_1}{2}+\frac{\eta
x_1}{2\sqrt{\eta^2-1}}+\frac{rx_1}{2\sqrt{\eta^2-1}}(1-2\eta(\eta+\sqrt{\eta^2-1}))\right ]\non
&\cdot& h_2(x_1,x_2,b_1,b_2) \cdot \alpha_s (t_2)\cdot
\exp\left [-S_{ab}(t_2) \right] \Bigr \}, \label{eq:A0qq}
\eeq
\beq
A_1(q^2)&=&8\pi m^2_{B_s} C_F\int dx_1 dx_2\int b_1 db_1 b_2 db_2
\phi_{B_s}(x_1,b_1)\phi^T_{D^*_s}(x_2,b_2)\cdot \frac{r}{1+r}\non
&\times & \Bigl \{2 [ 1+\eta-2 r x_2+r\eta x_2 ]\cdot
h_1(x_1,x_2,b_1,b_2) \cdot \alpha_s(t_1)\cdot \exp[-S_{ab}(t_1)]\non
& + & \left[2r_c+2 \eta r-x_1\right]\cdot h_2(x_1,x_2,b_1,b_2) \cdot
\alpha_s (t_2)\cdot \exp[-S_{ab}(t_2)] \Bigr \}, \label{eq:A1qq}
\eeq
\beq
A_2(q^2)&=&\frac{(1+r)^2(\eta-r)}{2r(\eta^2-1)}\cdot
A_1(q^2)- 8\pi m^2_{B_s} C_F\int dx_1 dx_2\int b_1 db_1 b_2
db_2\phi_{B_s}(x_1,b_1) \non & & \cdot \phi^L_{D^*_s}(x_2,b_2) \cdot
\frac{1+r}{\eta^2-1} \times \Bigl \{  \left[(1+\eta)(1-r)
-rx_2(1-2r+\eta(2+r-2\eta))\right]\non &&\cdot
h_1(x_1,x_2,b_1,b_2)\cdot \alpha_s(t_1) \cdot  \exp\left
[-S_{ab}(t_1) \right ] \non &+& \left[r+r_c(\eta-r)-\eta r^2+r
x_1\eta^2-\frac{x_1}{2}(\eta+r)+x_1 (\eta r
-\frac{1}{2})\sqrt{\eta^2-1}\right]\non &&\cdot h_2(x_1,x_2,b_1,b_2)
\cdot \alpha_s (t_2)\cdot \exp\left [-S_{ab}(t_2) \right] \Bigr \},
\label{eq:A2qq}
\eeq
where $r=m_{D^*_s}/m_{B_s}$, while $C_F$ and
$r_c$ is the same as in Eqs.(\ref{eq:f1q2},\ref{eq:f2q2}). And the
hard function $h(x_i,b_i)$, the hard scales $t_{1,2}$ and Sudakov
factor $S_{ab}$ are given in the Appendix \ref{sec:app1}.

For $\bar{B}_s^0\to D_s^{*+} l^-\bar\nu_l$ decays , the corresponding differential decay widths
can be written as ~\cite{prd79-054012}:
\beq
\frac{d\Gamma_L(\bar{B}_s^0 \to D_s^{*+} l^-
\bar{\nu}_l)}{dq^2}&=& \frac{G_F^2|V_{cb}|^2}{192 \pi^3  m_{B_s}^3}
\left ( 1-\frac{m_l^2}{q^2}\right )^2
\frac{\lambda^{1/2}(q^2)}{2q^2}\cdot \Bigg\{3m^2_l\lambda(q^2)A^2_0(q^2)\non
&& \hspace{-3cm}
+\frac{m^2_l+2q^2}{4m^2}\cdot \left [(m^2_{B_s}-m^2-q^2)(m_{B_s}+m)A_1(q^2)
 -\frac{\lambda(q^2)}{m_{B_s}+m}A_2(q^2) \right ]^2 \Bigg\},
\label{eq:dfds1}
\eeq
\beq
\frac{d\Gamma_\pm(\bar{B}_s^0 \to
D_s^{*+} l^-\bar\nu_l)}{dq^2}&=& \frac{G_F^2|V_{cb}|^2}{192 \pi^3
m_{B_s}^3}\left ( 1-\frac{m_l^2}{q^2}\right )^2
\frac{\lambda^{3/2}(q^2)}{2}\non & \times & \left \{
(m^2_l+2q^2)\left[\frac{V(q^2)}{m_{B_s}+m}\mp
\frac{(m_{B_s}+m)A_1(q^2)}{\sqrt{\lambda(q^2)}}\right]^2\right\},
\label{eq:dfds2}
 \eeq
  where $m=m_{D_s^*}$, and $\lambda(q^2) =
(m_{B_s}^2+m_{D_s^*}^2-q^2)^2 - 4 m_{B_s}^2 m_{D_s^*}^2$ is the
phase space factor. The combined transverse and total differential
decay widths are defined as:
 \beq
\frac{d\Gamma_T}{dq^2}=\frac{d\Gamma_+}{dq^2}
+\frac{d\Gamma_-}{dq^2} \; ,\quad
\frac{d\Gamma}{dq^2}=\frac{d\Gamma_L}{dq^2}
+\frac{d\Gamma_T}{dq^2} \; . \label{eq:dfdst} \eeq

\section{Numerical results and discussions} \label{sec:3}

In the numerical calculations we use the following input parameters
(here masses and decay constants are in units of
GeV)\cite{prd86-114025,pdg2012}
 \beq
    m_{D_s^+}&=&1.969, \quad m_{D_s^{*+}}=2.112,\quad m_{B_s}=5.367,
     \quad m_{c}=1.35 \pm 0.03, \non 
m_\tau &=& 1.777,\quad |V_{cb}|=(39.54 \pm 0.89)\times 10^{-3},
\quad \Lambda^{4}_{\overline{MS}} = 0.287,\non 
 f_{B_s}&=&0.24\pm0.02, \quad f_{D_s}=0.274,\quad \tau_{B_s^0}=1.497\times 10^{-12}s, \label{eq:inputs}
\eeq
while $f_{D_s^*}= f_{D_s} \sqrt{m_{D_s}/m_{D_s^*}}$.

\subsection{The form factors in the pQCD approach}

For the considered semileptonic decays, the differential decay rates strongly depend on the value and the shape
of the relevant form factors $F_{0,+}(q^2)$, $V(q^2)$ and $A_{0,1,2}(q^2)$.
The evaluation of these form factors play the key role in such works.
The two well-known traditional methods of evaluating the form factors are
the QCD sum rule for the low $q^2$ region and the Lattice QCD for the high $q^2$ region of $q^2\approx q_{max}^2$.

In the pQCD factorization approach \cite{li2003,prd67-054028,li1995}, one can also calculate the form
factors perturbatively in the lower $q^2$ region \cite{li2003}.
For B to light meson (such as $K, \pi, \rho, \etap, etc$) transitions, the values of
the relevant form factors have been evaluated successfully by employing the pQCD factorization approach for example in
Refs.\cite{li2003,prd85-094003,prd86-114025,prd78-014018,prd65-014007,prd63-074009,pqcd-ff}.
The pQCD predictions for the  form factors obtained in these papers agree very well with those obtained from the QCD sum rule.

For $B \to D^{(*)}$ transitions, one usually use the heavy quark effective theory (HQET) to
evaluate the form factors at the lower $q^2$ region, and consider the lattice QCD results at the higher $q^2$ region
with $q^2\approx q^2_{max}$.
In Ref.~\cite{fan2013a}, we evaluated the form factors for $B \to (D,D^*)$ transitions in the lower $q^2$ region
and obtained the pQCD predictions for the ratios $R(D)$ and $R(D^*)$ being consistent with those
measured by BaBar Collaboration.
In this paper, by using the expressions as given in Eqs.(\ref{eq:f1q2}-\ref{eq:f2q2},\ref{eq:Vqq}-\ref{eq:A2qq}) and the
definitions in Eq.~(\ref{eq:fpf0}), we calculate the values of the
form factors $F_0(q^2)$, $F_+(q^2)$, $V(q^2)$ and $A_{0,1,2}(q^2)$
for given value of $q^2$ in the lower $q^2$ region:  $0 \leq q^2 \leq m_\tau^2$.
For the form factors in the larger $q^2$ region, we make an extrapolation for them from
the lower $q^2$ region to larger $q^2$ region: $m_\tau^2 < q^2 \leq q_{max}^2 $
with $q_{max}^2=(m_{B_s}-m_{D_s})^2$ ( $q_{max}^2=(m_{B_s}-m_{D_s^*})^2$) for $B_s\to D_s$ ($B_s \to D_s^*$)  transition.
In this work we make the extrapolation by using the formula as given in
Ref.~\cite{prd79-054012}
 \beq
F(q^2)=\frac{F(0)}{1-a \frac{q^2}{m_{B_s}^2} + b(\frac{q^2}{m_{B_s}^2})^2},
\label{eq:fq2}
 \eeq
where $F$ stands for the form factors $F_{0,+}, V, A_{0,1,2}$.
The parameters $a$ and $b$ in above equation are determined
by the fitting to the pQCD predicted values of the form factors at the sixteen points in
the lower $q^2$ region, as illustrated explicitly in Fig.2 for the case of $F_{0,+}(q^2)$.

In Table \ref{tab:ffab}, we list the pQCD predictions
for the form factors $F_{0,+}(q^2), V(q^2)$ and $A_{0,1,2}(q^2)$ and the corresponding
parametrization constants ``a" and ``b" in Eq.~(\ref{eq:fq2}) for
$\bar{B}_{s}^{0} \to (D_s,D_s^*)$ transitions at the scale $q^2=0$ and $q^2=m_\tau^2$.
The theoretical error of the form factors as shown in
Table \ref{tab:ffab} is the total error: a combination of the three
major theoretical errors from the uncertainties of the parameter
$\omega_{B_s}=0.50\pm 0.05$ GeV, $f_{B_s}=0.24\pm 0.02$ GeV and $m_c=1.35\pm 0.03$ GeV.
For the parametrization constants ``a" and ``b", they do not depend on the variation of $f_{B_s}$, and
the total errors are a combination of the two theoretical errors from the uncertainties of
$\omega_{B_s}$  and $m_c$.

\begin{table}[thb]
\begin{center}
\caption{The pQCD predictions for the form factors $F_{0,+}(q^2), V(q^2)$ and
$A_{0,1,2}(q^2)$ at the scale $q^2=0$ and $q^2=m_\tau^2$. The parametrization constants ``a" and ``b"
are also listed in last two columns.}
\label{tab:ffab}\vspace{0.2cm}
\begin{tabular}{l| cc |c c} \hline \hline
      ~~~~\quad& $F(0)$ &  $F(m_\tau^2)$ & $a$ & $b$ \\ \hline\hline
$F_0^{\bar{B}_{s}^{0}\to D_{s}}$ &
$0.55^{+0.15}_{-0.12}$& $0.67^{+0.17}_{-0.13}$ &$1.69^{+0.06}_{-0.10}$&$0.78^{+0.23}_{-0.33}$ \\ \hline
$F_+^{\bar{B}_{s}^{0}\to D_{s}}$ &
$0.55^{+0.15}_{-0.12}$ & $0.74^{+0.19}_{-0.15}$ &$2.44^{+0.05}_{-0.08}$& $1.70^{+0.18}_{-0.33}$ \\ \hline \hline
$V^{\bar{B}_{s}^{0}\to D_{s}^*}$ &
$0.62^{+0.15}_{-0.12}$ &$0.83^{+0.18}_{-0.15}$&$2.48^{+0.12}_{-0.16}$& $1.66^{+0.12}_{-0.12}$ \\ \hline
$A_0^{\bar{B}_{s}^{0}\to D_{s}^*}$ &
$0.47^{+0.11}_{-0.09}$ &$0.63^{+0.13}_{-0.11}$&$2.49^{+0.09}_{-0.12}$& $1.74^{+0.06}_{-0.00}$ \\ \hline
$A_1^{\bar{B}_{s}^{0}\to D_{s}^*}$ &
$0.49^{+0.12}_{-0.10}$ &$0.60^{+0.13}_{-0.11}$&$1.64^{+0.09}_{-0.15}$& $0.59^{+0.06}_{-0.13}$ \\ \hline
$A_2^{\bar{B}_{s}^{0}\to D_{s}^*}$ &
$0.52^{+0.13}_{-0.10}$ &$0.67^{+0.15}_{-0.12}$&$2.33^{+0.09}_{-0.16}$& $1.81^{+0.00}_{-0.20}$ \\
\hline \hline
\end{tabular}
\end{center} \end{table}


\begin{figure}[thb]
\begin{center}
\vspace{-0.5cm} \centerline{\epsfxsize=8cm\epsffile{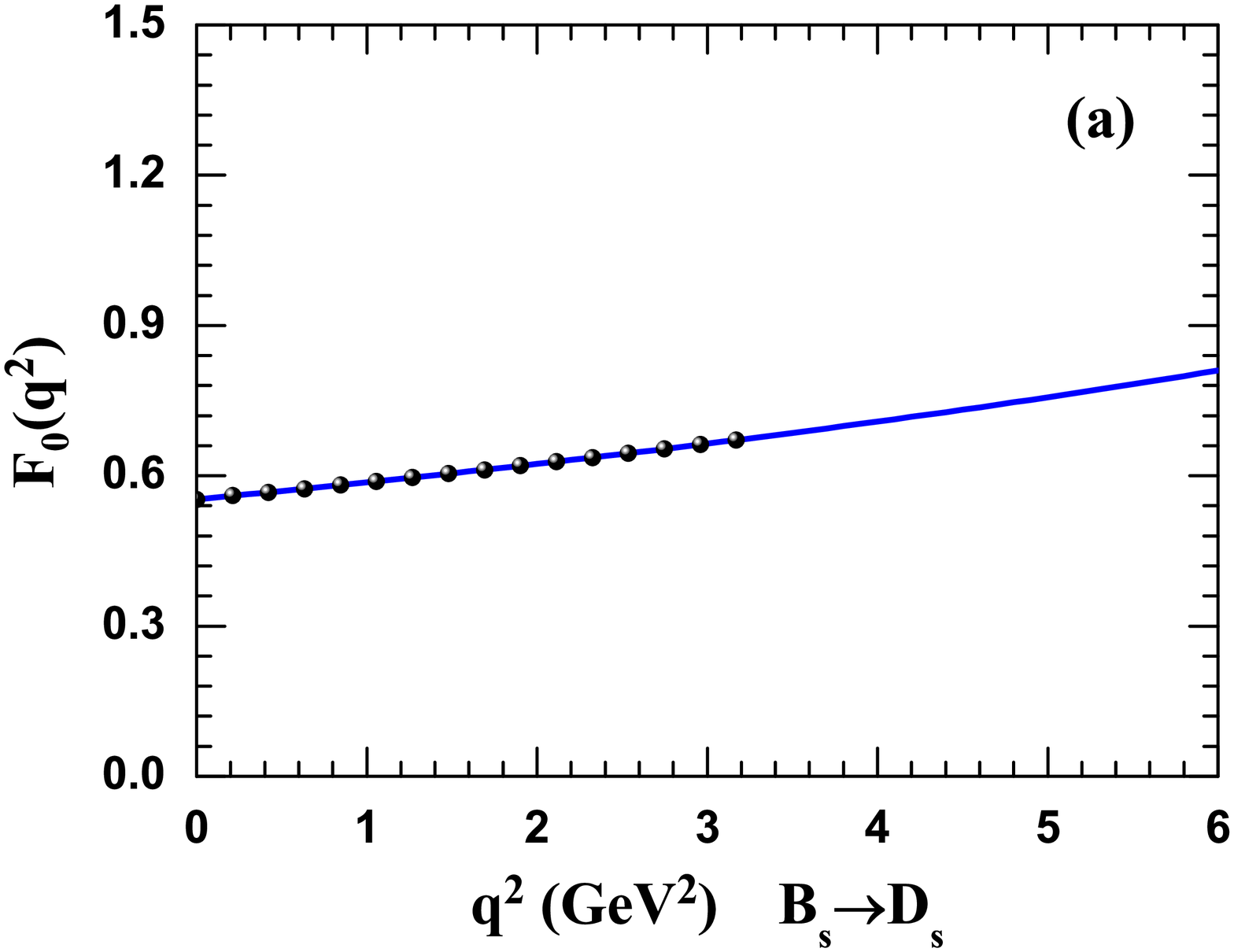}
\hspace{-0.8cm}\epsfxsize=8cm\epsffile{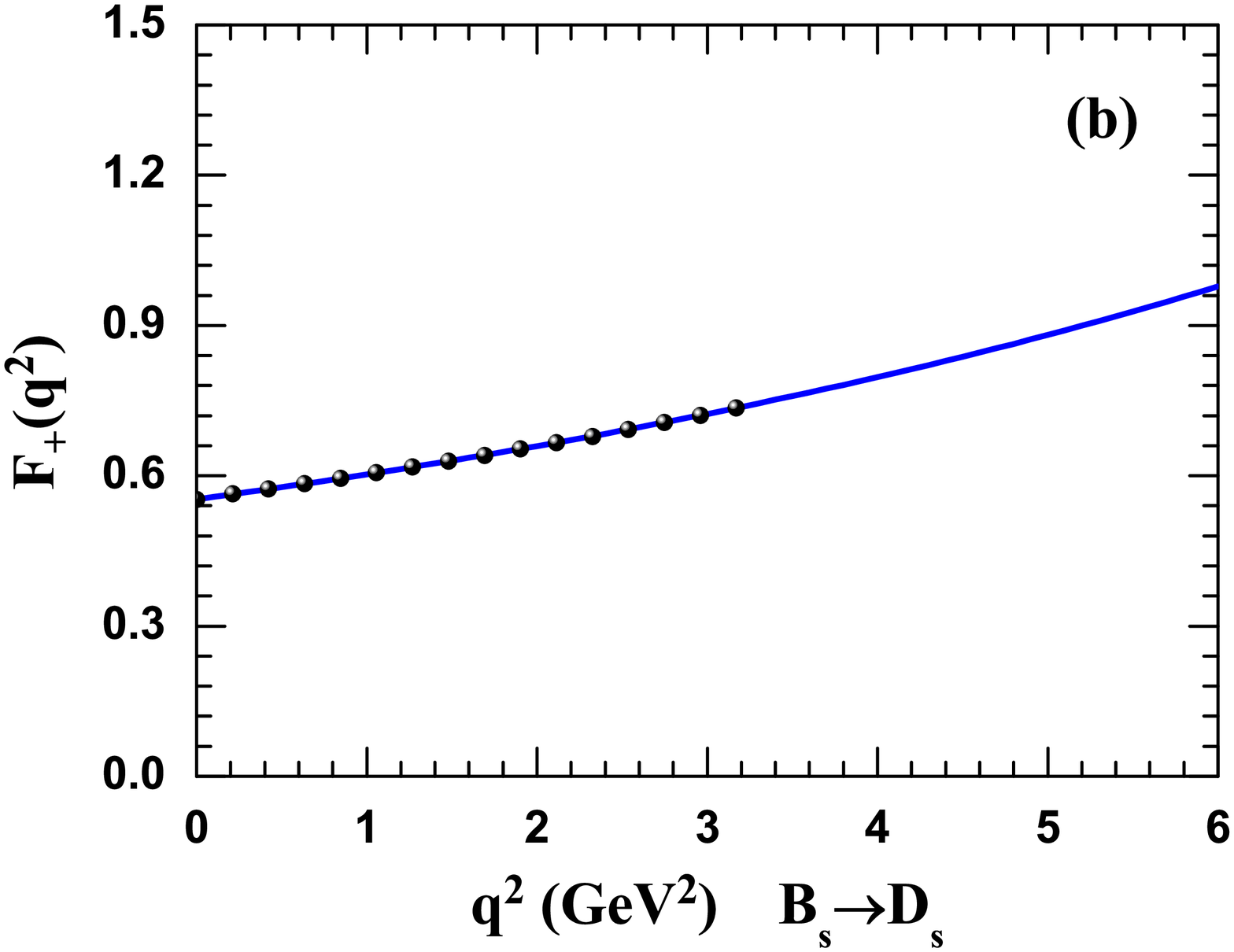}}
\caption{ The $q^2$-dependence of the form factors $F_{0,+}(q^2)$ in the pQCD approach for the case of
$B_s \to D_s$ transition. The dots in (a) and (b) refer to the pQCD predictions
for each given value of $q^2$ in the range of $0 \leq q^2 \leq
m_\tau^2$. } \label{fig:fig2}
\end{center}
\end{figure}

\subsection{Differential decay widths and branching ratios }

From the differential decay rates as given in
Eqs.~(\ref{eq:dg1},\ref{eq:dg1b},\ref{eq:dfds1}-\ref{eq:dfdst}), it
is straightforward to make the integration over the range of $m_l^2
\leq q^2 \leq (m_{B_s}-m)^2$ with $m=(m_{D_s},m_{D_s^*})$. The pQCD
predictions for the branching ratios of the semileptonic decays
$\bar{B}^0_s \to D_s^{(+,*+)} \tau^- \bar{\nu}_\tau$ and
$\bar{B}_s^0 \to D_s^{(+,*+)} l^- \bar{\nu}_l$ are the following:
\beq
{\cal B}(\bar{B}_{s}^{0} \to D_{s}^{+} \tau^- \bar{\nu}_\tau)
&=& (0.84^{+0.34}_{-0.24}(\omega_{B_{s}})^{+0.15}_{-0.13}
(f_{B_{s}})^{+0.04}_{-0.04}(V_{cb})^{+0.03}_{-0.00}(m_c))\%, \non
{\cal B}(\bar{B}_{s}^{0} \to D_{s}^{+} l^- \bar{\nu}_l) &=&
(2.13^{+1.05}_{-0.68}(\omega_{B_{s}})^{+0.37}_{-0.34}
(f_{B_{s}})^{+0.10}_{-0.10}(V_{cb})^{+0.07}_{-0.03}(m_c))\%,
\label{eq:br11}
\eeq
\beq
   {\cal B}(\bar{B}_{s}^{0}\to D_{s}^{*+} \tau^- \bar{\nu}_\tau) &=&
(1.44^{+0.43}_{-0.34}(\omega_{B_{s}})^{+0.25}_{-0.23}
(f_{B_{s}})^{+0.07}_{-0.06}(V_{cb})^{+0.07}_{-0.07}(m_c))\%, \non
{\cal B}(\bar{B}_{s}^{0}\to D_{s}^{*+} l^- \bar{\nu}_l) &=&
(4.76^{+1.65}_{-1.25}(\omega_{B_{s}})^{+0.83}_{-0.76}
(f_{B_{s}})^{+0.22}_{-0.21}(V_{cb})^{+0.18}_{-0.17}(m_c))\%,
 \label{eq:br12}
 \eeq
where the four major theoretical errors come
from the uncertainties of the input parameters $\omega_{B_s}=0.50\pm
0.05$ GeV, $f_{B_s}=0.24\pm 0.02$ GeV, $|V_{cb}|=(39.54 \pm 0.89)
\times 10^{-3}$ and $m_c=1.35\pm 0.03$ GeV.

\begin{figure}[thb]
\begin{center}
\vspace{-0.5cm} \centerline{\epsfxsize=6cm\epsffile{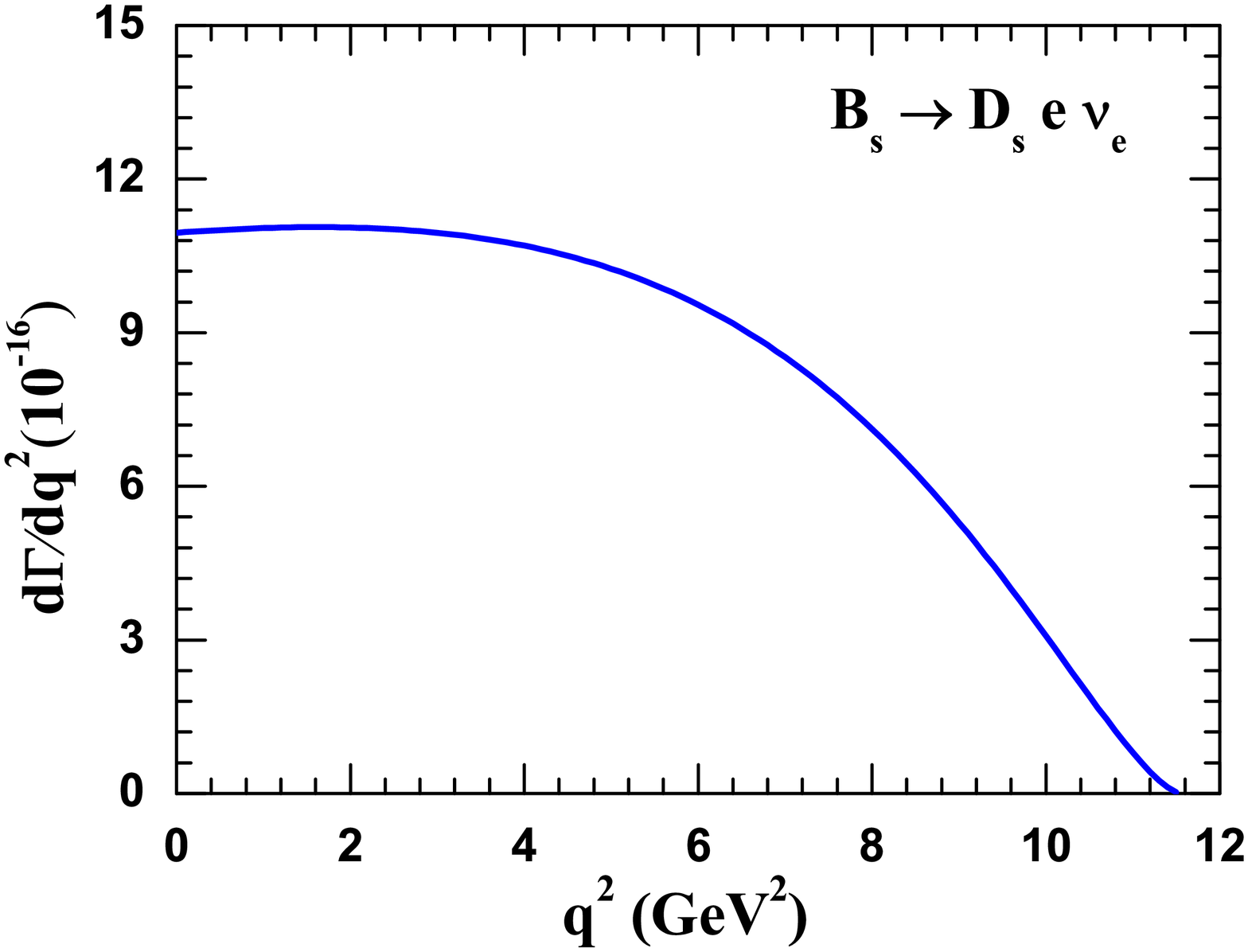}
\hspace{-0.5cm}\epsfxsize=6cm\epsffile{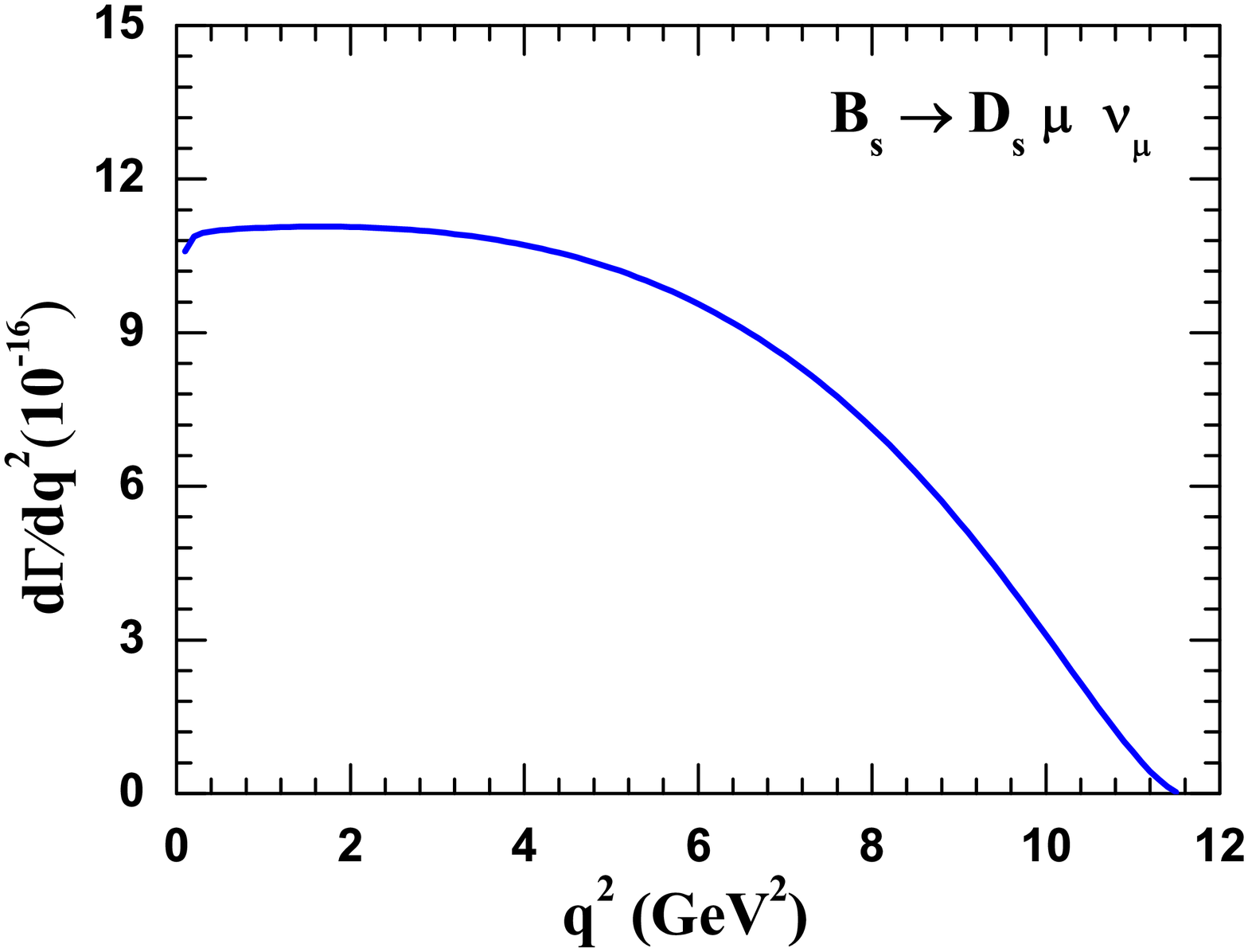}
\hspace{-0.5cm}\epsfxsize=6cm\epsffile{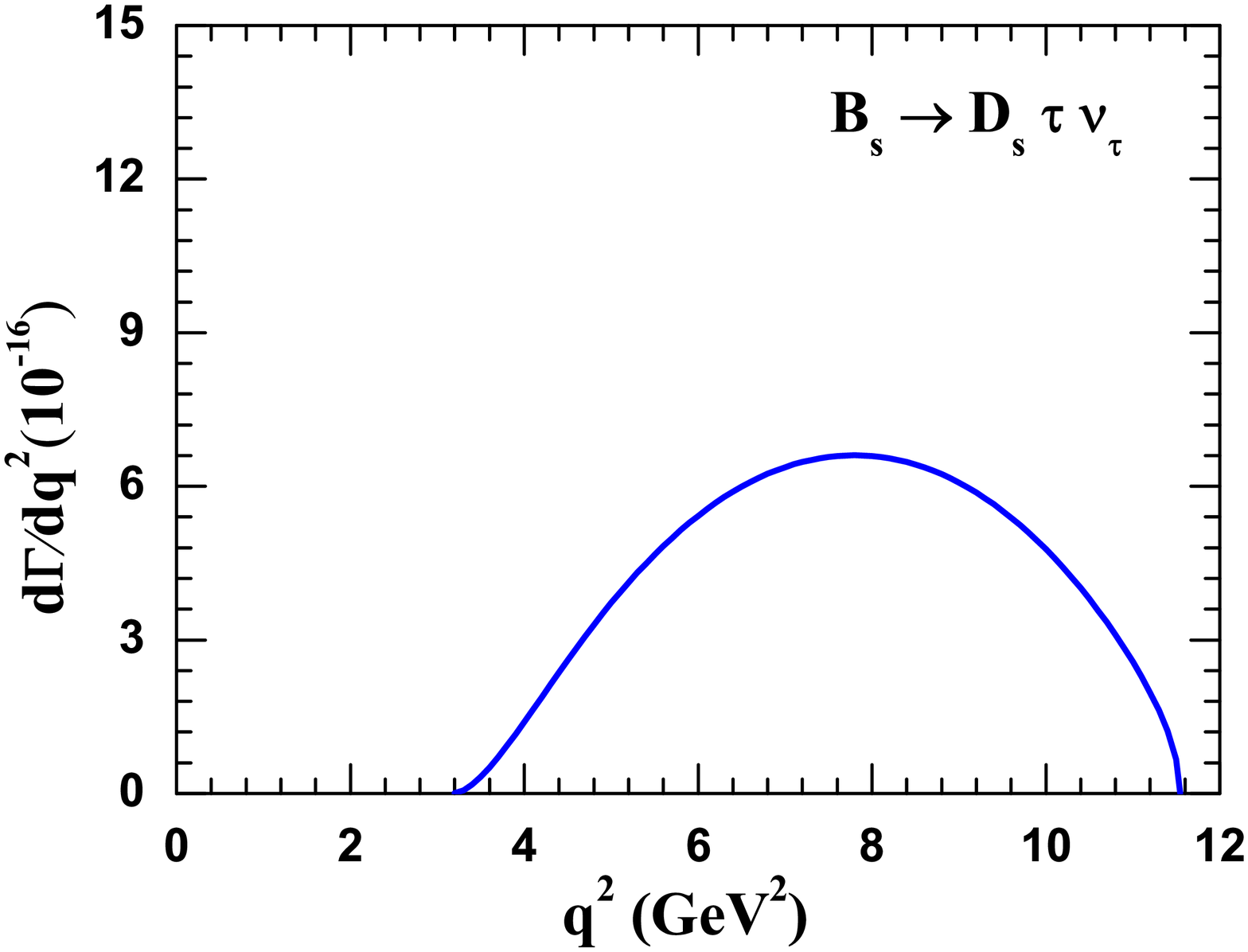} } \caption{The
theoretical predictions for the $q^2$-dependence of $d\Gamma/dq^2$
for the semileptonic decays $B_s \to D_s^+l^-\bar{\nu}_l $ with
$l=e,\mu,\tau$ in the pQCD approach. Here $q^2_{max}$ = 11.55
GeV$^2$.} \label{fig:fig3}
\end{center}
\end{figure}

\begin{figure}[thb]
\begin{center}
\vspace{-1cm} \centerline{\epsfxsize=6cm\epsffile{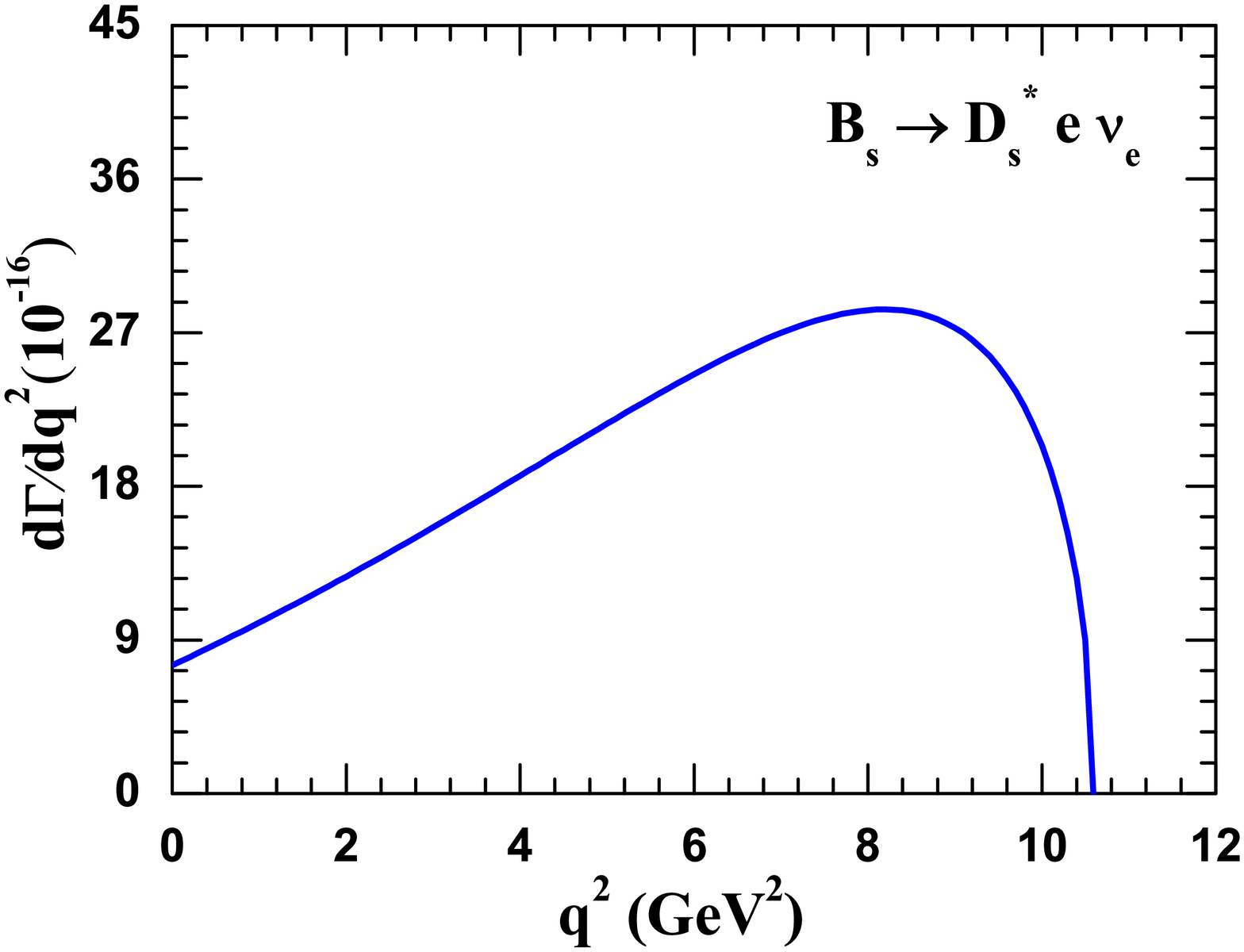}
\hspace{-0.5cm}\epsfxsize=6cm\epsffile{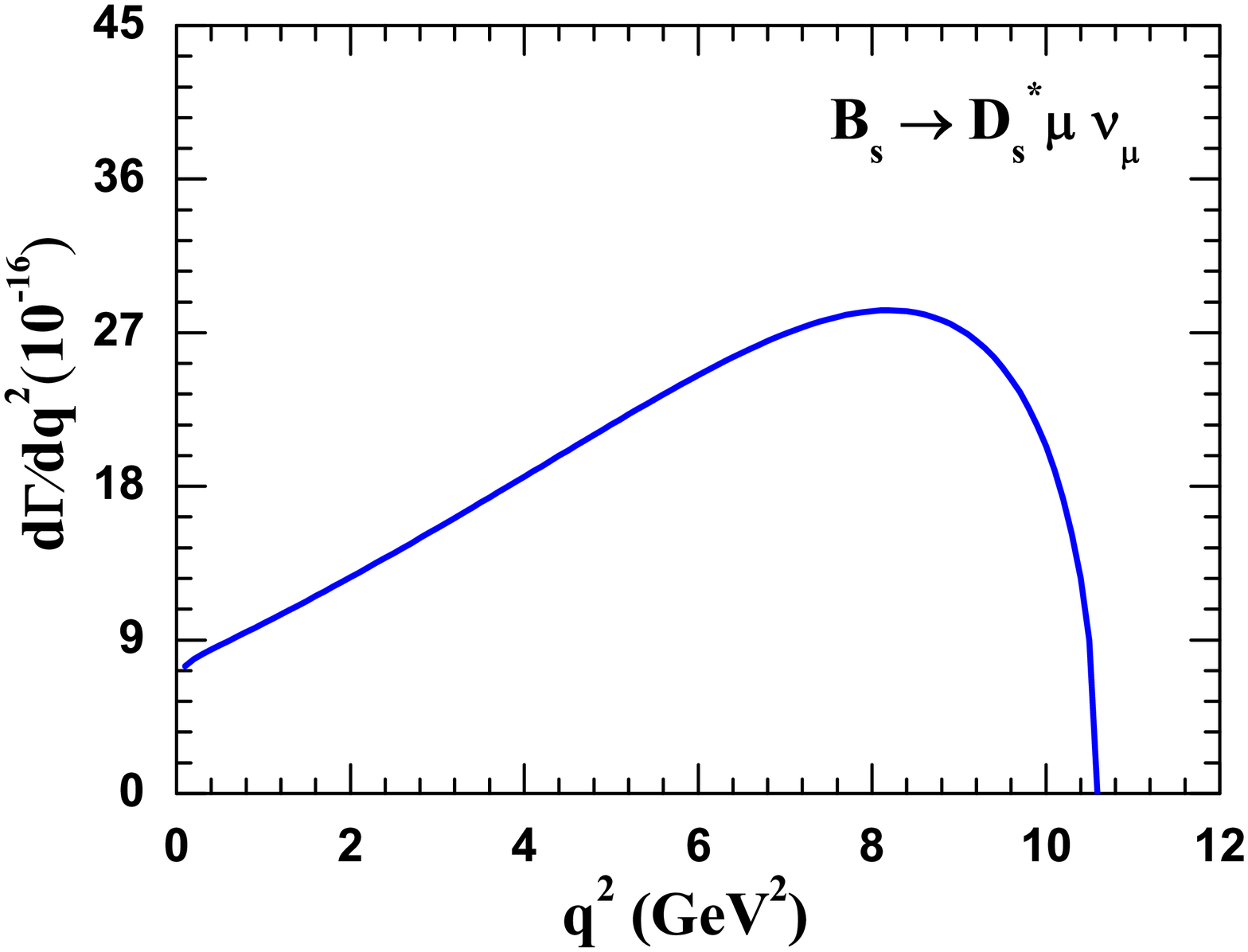}
\hspace{-0.5cm}\epsfxsize=6cm\epsffile{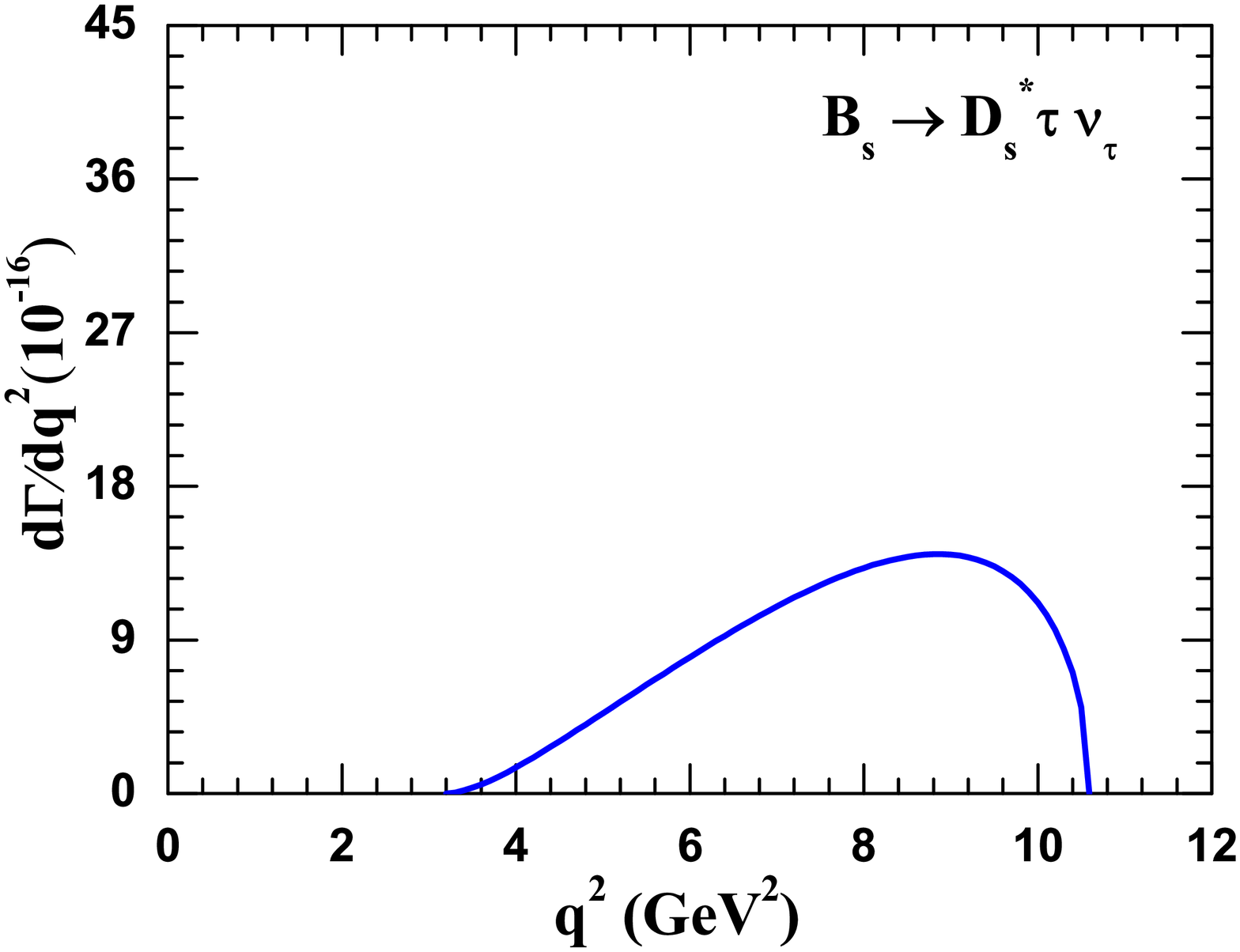}  }
\caption{The theoretical predictions for the $q^2$-dependence of
$d\Gamma/dq^2$ for the semileptonic decays $B_s \to
D_s^{*+}l^-\bar{\nu}_l $ with $l=e,\mu,\tau$ in the pQCD approach.
Here $q^2_{max}$ = 10.60 GeV$^2$.} \label{fig:fig4}
\end{center}
\end{figure}

In Figs.~\ref{fig:fig3} and \ref{fig:fig4}, we show the
$q^2$-dependence of the theoretical predictions for the differential
decay rates $d\Gamma/dq^2$ for $\bar{B}_s^0\to D_s^{+} l^-
\bar{\nu}_l$ and $\bar{B}_s^0\to D_s^{*+} l^- \bar{\nu}_l$ decays
calculated by using  the pQCD factorization approach.

In Table \ref{tab:br31}, the pQCD predictions for the branching
ratios of the considered decay modes are listed in column two where the
theoretical errors from different sources have been added in
quadrature. As a comparison, we also show the theoretical predictions
as given previously in Refs.~\cite{epjc51-601,prd78-054011,prd80-014005,prd82-094031,jpg39-045002,prd87-034033}.
One can see from the numerical results as shown in Table \ref{tab:br31} that:
\begin{enumerate}
\item[(i)]
For the branching ratios, although the previous theoretical predictions basically agree within a factor of 2,
they are rather different from each other. The reason is that these results were obtained by using different
methods to evaluate the relevant form factors.
The pQCD predictions, however,  agree well with previous results within one standard deviation.

\item[(ii)]
The theoretical errors of the pQCD predictions for the branching ratios are still large,
say $\sim 35\%$. It is therefore necessary to define the ratios $R(X)$
among the branching ratios of the individual decays, since the theoretical errors are greatly canceled in these ratios.

\end{enumerate}

\begin{table}[thb]
\begin{center}
\caption{ The pQCD predictions for the branching ratios (in units of $10^{-2}$) of
the considered decay modes. The theoretical predictions are given in
Refs.~\cite{epjc51-601,prd78-054011,prd80-014005,prd82-094031,jpg39-045002,prd87-034033}
are listed as a comparison.}
\label{tab:br31}\vspace{0.2cm}
\begin{tabular}{l |c |c c  c c c} \hline \hline
 Channel & pQCD & CQM\cite{epjc51-601}& QCDSRs\cite{prd78-054011}
& \cite{prd80-014005,prd82-094031} & IAMF\cite{jpg39-045002} & RQM\cite{prd87-034033} \\
\hline 
$\bar{B}_{s}^0 \to D_{s}^+ \tau^- \bar{\nu}_\tau$ &
$0.84^{+0.38}_{-0.28}$ & $--$ & $--$ & $0.33^{+0.14}_{-0.11}$& $0.47-0.55$& $0.62 \pm 0.05$\\
$\bar{B}_{s}^0 \to D_{s}^+ l^- \bar{\nu}_l$ &
$2.13^{+1.12}_{-0.77}$ &$2.73-3.00$ & $2.8-3.8$ & $1.0^{+0.4}_{-0.3}$& $1.4-1.7$& $2.1 \pm 0.2$\\
\hline 
$\bar{B}_{s}^0 \to D_{s}^{*+}\tau^-\bar{\nu}_\tau$ &
$1.44^{+0.51}_{-0.42}$ &$--$& $--$ & $1.3^{+0.2}_{-0.1}$& $1.2-1.3$& $1.3 \pm 0.1$\\
$\bar{B}_{s}^0\to D_{s}^{*+} l^- \bar{\nu}_l$ &
$4.76^{+1.87}_{-1.49}$ &$7.49-7.66$& $1.89-6.61$ & $5.2 \pm 0.6$& $5.1-5.8$& $5.3 \pm 0.5$\\
\hline \hline
\end{tabular}
\end{center} \end{table}


\subsection{The pQCD predictions for $R(X)$-ratios }

It is straightforward to calculate the four $R(X)$ ratios of the branching ratios
for $\bar{B}^0_s\to D_s^{(*)}l \bar{\nu}_l$ decays by using the definitions
as made previously in Eqs.~(5,6), the corresponding pQCD predictions are listed
in Table \ref{tab:ratios}. As a comparison, the pQCD predictions for the
corresponding $R(X)$ ratios for $B\to D^{(*)}l \bar{\nu}_l$ decays as given
in Ref.~\cite{fan2013a} are also shown in Table \ref{tab:ratios}.
The errors of the pQCD predictions as given in Table \ref{tab:ratios} are the combination
of the major theoretical errors come from the uncertainties of
$\omega_{B_s}=0.50\pm 0.05$ GeV ($\omega_{B}=0.40\pm 0.04$ GeV)
and $m_c=1.35\pm 0.03$ GeV, while those induced
by the variations of $f_{B_s}$ ($f_{B}$) and $|V_{cb}|$ are canceled
completely in the pQCD predictions for the $R(X)$-ratios.

\begin{table}[thb]
\begin{center}
\caption{The pQCD predictions for the eight $R(X)$ ratios for both
$\bar{B}^0_s\to D_s^{(*)}l \bar{\nu}_l$
and $B\to D^{(*)} l \bar{\nu}_l$ \cite{fan2013a} decays.
Currently available measurements are BaBar results \cite{prl109-101802}
as shown in Eq.~(1).}
\label{tab:ratios}\vspace{0.2cm}
\begin{tabular}{ cc |c c} \hline \hline
$R(D_s)$& $R(D_s^*)$ &  $R_{D_s}^l$& $R_{D_s}^\tau $  \\ \hline
$0.392\pm 0.022$ & $0.302\pm 0.011$ & $0.448^{+0.058}_{-0.041}$
& $0.582^{+0.071}_{-0.045}$\\ \hline \hline
$R(D)$& $R(D^*)$ &  $R_D^l$& $R_D^\tau$  \\ \hline
$0.430^{+0.021}_{-0.026}$& $0.301\pm 0.013$& $0.450^{+0.064}_{-0.051}$& $0.642^{+0.081}_{-0.070}$
\\ \hline \hline
\end{tabular}
\end{center} \end{table}


From the numerical results as listed in Table \ref{tab:ratios}
one can see the following points:
\begin{enumerate}
\item[(i)]
The theoretical errors of the pQCD predictions for the R(X) ratios are less than
$13\%$, much smaller than those for the branching ratios.
These ratios could be measured at the forthcoming Super-B experiments.

\item[(ii)]
The ratio $R(D_{s})$ and $R(D_{s}^{*})$ are defined here in the same way as
the ratios $R(D^{(*)})$ in Refs.~\cite{prl109-101802,prd85-094025}.
These ratios generally measure the mass effects of heavy $m_\tau$ against the
light $m_e$ or $m_\mu$.

\item[(iii)]
The new ratio $R_{D_{s}}^l$ and $R_{D_{s}}^\tau$ will measure the
effects induced by the variations of the form factors for
$\bar{B}_s^0 \to D_s$ and $\bar{B}_s^0 \to D_s^*$ transitions.
In other words, the new ratios $R_{D_{s}}^{l,\tau}$
may be more sensitive to the QCD dynamics which
controls the $\bar{B}_s^0 \to D_s^{(*)}$ transitions than the
ratios $R(D_s^{(*)})$.

\item[(iv)]
In the limit of $SU(3)_F$ flavor symmetry, the four ratios defined for
$\bar{B}^0_s\to D_s^{(*)}l \bar{\nu}_l$ decays should be very similar with the
corresponding ones for $B\to D^{(*)} l \bar{\nu}_l$ decays.
The pQCD predictions as listed in Table \ref{tab:ratios} do support this
expectation. The breaking of $SU(3)_F$ flavor symmetry  is less than $10\%$.

\item[(v)]
At present, only the ratio $R(D)$ and $R(D^*)$ have been measured  by
Belle and BaBar \cite{prl109-101802,babar-2008,belle-2007,belle-2010,U2013}.
In order to check if the BaBar's anomaly of $R(D^{(*)})$ do exist or not for
$\bar{B}^0_s\to D_s^{(*)}l \bar{\nu}_l$ decays, and to test the
$SU(3)_F$ flavor symmetry among $\bar{B}^0_s\to D_s^{(*)}l \bar{\nu}_l$ and
$B\to D^{(*)} l \bar{\nu}_l$ decays, we strongly suggest the forthcoming
Super-B experiments to measure these four new ratios $R(D_s^{(*)})$ and $R_{D_{s}}^{l,\tau}$.

\end{enumerate}

\section{Summary and Conclusions} \label{sec:summary}

In this paper, we studied the semileptonic decays $\bar{B}_s^0 \to D_s^{(*)+} l^- \bar{\nu}_l$
in the framework of the SM by employing the pQCD factorization approach.
We calculated the branching ratios ${\cal B}(\bar{B}_s^0 \to D_s^{(*)+} l^-
\bar{\nu}_l)$ and the four ratios of the branching ratios: $R(D_s)$, $R(D_s^{*})$, $R_{D_{s}}^l$
and $R_{D_{s}}^\tau$. From the numerical results and phenomenological analysis we found that
\begin{enumerate}
\item[(i)]
For the branching ratios $Br(\bar{B}_s^0 \to D_s^{(*)+} l^- \bar{\nu}_l)$, the pQCD predictions generally
agree well with previous results obtained by using different methods to evaluate the relevant form factors.

\item[(ii)]
For the four new ratios $R(D_s^{(*)})$ and $R_{D_{s}}^{l,\tau}$, the pQCD predictions are
\beq
R(D_s) &=& 0.392 \pm 0.022, \quad R(D_s^{*})= 0.302 \pm 0.011,\non
R_{D_{s}}^l&=&0.448^{+0.058}_{-0.041}, \quad  R_{D_{s}}^\tau=0.582^{+0.071}_{-0.045}.
\eeq

\item[(iii)]
The pQCD predictions do support the $SU(3)_F$ flavor symmetry between $\bar{B}_s^0 \to D_s^{(*)+} l^- \bar{\nu}_l$
and $B \to D^{(*)} l^- \bar{\nu}_l$ decay modes, while the breaking of $SU(3)_F$ flavor symmetry  is less
than $10\%$.

\item[(iv)]
Based on our analysis, we strongly suggest the forthcoming Super-B experiments to measure the new ratios
$R(D_s^{(*)})$  and $R_{D_{s}}^{l,\tau}$.

\end{enumerate}

\begin{acknowledgments}
We wish to thank Cai-Dian L\"u and Xin Liu for valuable discussions.
This work was supported by the National Natural Science Foundation of
China under Grant No.10975074 and 11235005.

\end{acknowledgments}


\appendix

\section{Relevant functions}\label{sec:app1}

The threshold resummation factor $S_t(x)$ is adopted from \cite{prd65-014007}:
\beq
\label{eq-def-stx} S_t=\frac{2^{1+2c}\Gamma(3/2+c)}{\sqrt{\pi}\Gamma(1+c)}[x(1-x)]^c,
\eeq
and we here set the  parameter $c=0.3$. The hard functions
$h_1$ and $h_2$ come form the Fourier transform and can be written as
\beq
\begin{aligned}
h_1(x_1,x_2,b_1,b_2)&=K_0(\beta_1 b_1)
[\theta(b_1-b_2)I_0(\alpha_1 b_2)K_0(\alpha_1 b_1)\\
&+\theta(b_2-b_1)I_0(\alpha_1 b_1)K_0(\alpha_1 b_2)]S_t(x_2),
\end{aligned}
\eeq
\beq
\begin{aligned}
h_2(x_1,x_2,b_1,b_2)&=K_0(\beta_2 b_1)
[\theta(b_1-b_2)I_0(\alpha_2 b_2)K_0(\alpha_2 b_1)\\
&+\theta(b_2-b_1)I_0(\alpha_2 b_1)K_0(\alpha_2 b_2)]S_t(x_2),
\end{aligned}
\eeq
where $K_0$ and $I_0$ are modified Bessel functions, and
\beq
\alpha_1 = m_{B_s}\sqrt{x_2r \eta^+},\quad
\alpha_2=m_{B_s}\sqrt{x_1 r \eta^+ - r^2+r_c^2},\quad
\beta_1 = \beta_2=m_{B_s}\sqrt{x_1x_2 r \eta^+},\quad
\eeq
where $r=m_{D^{(*)}}/m_{B_s}, r_c=m_c/m_{B_s}$.

The factor $\exp[-S_{ab}(t)]$ contains the Sudakov logarithmic
corrections and the renormalization group evolution effects of both
the wave functions and the hard scattering amplitude with
$S_{ab}(t)=S_B(t)+S_M(t)$ \cite{prd65-014007,prd63-074009},
\beq
S_B(t)&=&s\left(x_1\frac{m_{B_s}}{\sqrt{2}},b_1\right)
+\frac{5}{3}\int_{1/b_1}^{t}\frac{d\bar{\mu}}{\bar{\mu}}\gamma_q(\alpha_s(\bar{\mu})),\\
S_M(t)&=&s\left (x_2\frac{m_{B_s}}{\sqrt{2}} r\eta^+,b_2 \right )
+s\left ((1-x_2)\frac{m_{B_s}}{\sqrt{2}} r \eta^+,b_2 \right)
+2\int_{1/b_2}^{t}\frac{d\bar{\mu}}{\bar{\mu}}
\gamma_q(\alpha_s(\bar{\mu})),
\eeq
with the quark anomalous dimension $\gamma_q=-\alpha_s/\pi$.
The explicit expressions of the functions $s(Q,b)$ can be found for example in Appendix A of
Ref.~\cite{prd63-074009}. The hard scales $t_i$ in
Eqs.(\ref{eq:f2q2}-\ref{eq:A2qq}) are chosen as the largest scale of
the virtuality of the internal particles in the hard $b$-quark decay diagram,
\beq
t_1&=&\max\{m_{B_s}\sqrt{x_2 r \eta^+}, 1/b_1, 1/b_2\},\non
t_2&=&\max\{m_{B_s}\sqrt{x_1 r \eta^+ - r^2+r_c^2},1/b_1, 1/b_2\}.
\eeq


\end{document}